# Modified Newton's rings: II

T Sai Chaitanya, Rajiv kumar, V Sai Krishna, B Shankar Anandh and K S Umesh[#]

Department of Physics, Sri Sathya Sai University,
Brindavan Campus, P.O. Kadugodi,
Bangalore, 560067, India.

# Author for correspondence: umeshks@sssu.edu.in

## ABSTRACT

In an earlier work[3] reported by our group, a version of Newton's rings experiment called Modified Newton's rings was proposed. The present work is an extension of this work. Here, a general formula for wavelength has been derived, applicable for a plane of observation at any distance. A relation between the focal length and the radius curvature is also derived for a plano-convex lens which is essentially used as a concave mirror. Tracker, a video analysis software, freely downloadable from the net, is employed to analyze the fringes captured using a CCD camera. Two beams which give rise to interference fringes in conventional Newton's rings and in the present setup are clearly distinguished.

Submitted to: European Journal of Physics

## 1. Introduction

Recently, a novel method has been proposed and setup in our undergraduate laboratory to determine the wavelength of laser light using modified Newton's rings. These rings are similar, yet distinctly different from conventional Newton's rings. For instance, these fringes are essentially Fizeau[1] fringes like Newton's rings, but the order is maximum for central disc unlike that in Newton's rings[1,2]. Detailed comparison of these two is provided in the reference[3], along with its principle, experimental details, coherence requirements etc. In fact, this experiment can easily be adopted in the undergraduate laboratory due to its simplicity. In the literature[4], there is an elegant article on classroom demonstration of Newton's rings where the authors derive the formula for wavelength for the case when the convex surface of the lens is illuminated from a point source.

## 2. Principle

The schematic diagram of the experimental setup is shown in the figure 1. Laser beam, expanded using a microscope objective, is collimated using a lens. The collimated beam is incident on the plano-convex lens after passing through a beam splitter BS, inclined to the beam at an angle of $45^o$. There are two reflected beams. One is from the plane surface: it retraces the path upto the BS, gets reflected again at the BS which remains parallel and the second is from the curved surface of plano-convex lens which is a converging beam. These two beams, on superposition, give rise to interference fringes similar in appearance to Newton's rings. These fringes are directed by the beam splitter to the observation plane placed in a perpendicular direction, convenient for measurements. Fringe diameters at the unit magnification plane (UMP) can be measured using a travelling microscope. A unit magnification plane is one in which the ring diameters are identical to those at the plano-convex lens.

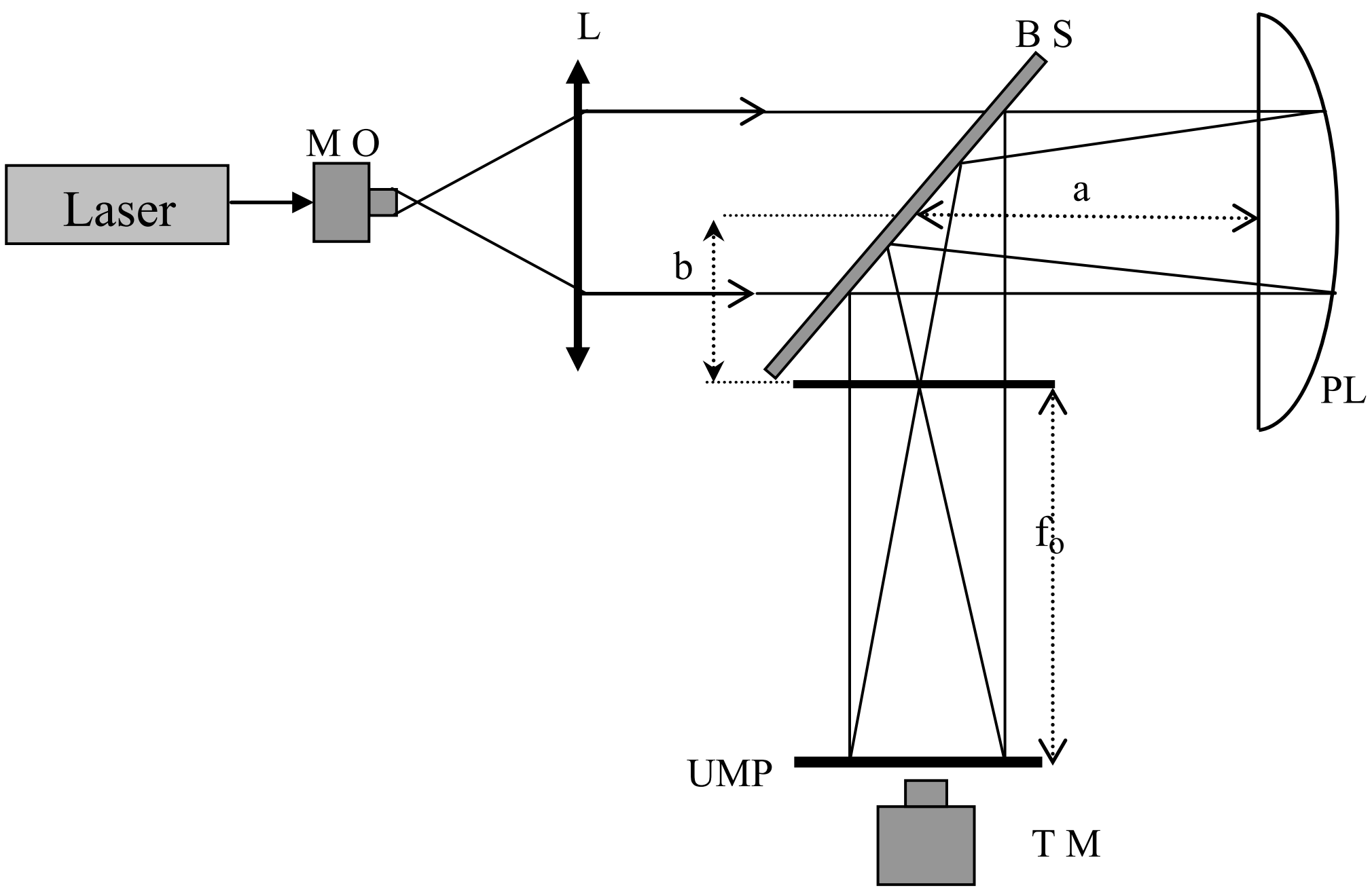

Fig.(1) : Schematic diagram for Modified Newton's rings: $a + b = f_o$ , MO: Microscope objective, L: Collimating lens, BS: beam splitter, PL: plano-convex lens, UMP: Unit Magnification Plane, TM: travelling microscope.

Placing the observation screen at the UMP is critical to the experiment since the fringes are not localized and thus fringes are formed wherever the two beams overlap. The UMP can be located as follows. As shown in figure 1, the reflected beams from the plane and curved surfaces of the lens, can be easily distinguished. When an observation plane is placed at an arbitrary distance, one can see a spot of light on circular beam of constant diameter. The rays from the plane surface, remain collimated unlike the ones from the curved surface which focus at a distance $f_o$, from the lens. The UMP lies exactly at a distance $f_o$, from the focal point of the beam reflected from the curved surface, that is, at a distance $2f_o$, from the plano-convex lens, PL along the path of the beam. In the figure 1, a+b is approximated to be equal to $f_o$ , neglecting the thickness of the lens compared to its focal length.

The formula[3] that is used in this method when the measurements are made in the UMP, is given by

$$\lambda = \frac{(r_m^2 - r_{m+p}^2)n}{pR} \qquad \text{----- (1)}$$

where R is radius of curvature of plano-convex lens, n is the refractive index of the lens, $r_{m+p}$ and $r_m$ are the radii of the rings of respective orders. Interference fringes in the form of rings, have their order maximum for the central disc and it reduces as we move to the periphery. This is exactly opposite to that in Newton's rings. Using a travelling microscope, the radii $r_m$ of various rings can be measured. Plot of m versus $r_m^2$ gives a linear graph whose slope is given by $\frac{(r_m^2 - r_{m+p}^2)}{p}$. The radius of curvature of the plano-convex lens is generally measured using a spherometer. Thus the wavelength can be determined by knowing the refractive index of plano-convex lens which is 1.5 for glass.

## 3. Derivation of Relation between R and $f_o$:

In this experiment, the determination of R is simplified since we are already able to measure $f_o$ to locate the unit magnification plane as explained earlier. Therefore, if we know the relation between R and $f_o$, the analysis is simplified, it also improves the accuracy of measurement as well. As can be seen from the schematic diagram in figure 2, in this particular setup the plano-convex lens is not used as a lens but rather as a concave mirror. Hence, we need a formula that relates the focal length, $f_0$ of the concave mirror and radius of curvature of the lens, R.

As it will be shown below, the formula turns out to be

$$R = 2nf_o \qquad \text{----- (2)}$$

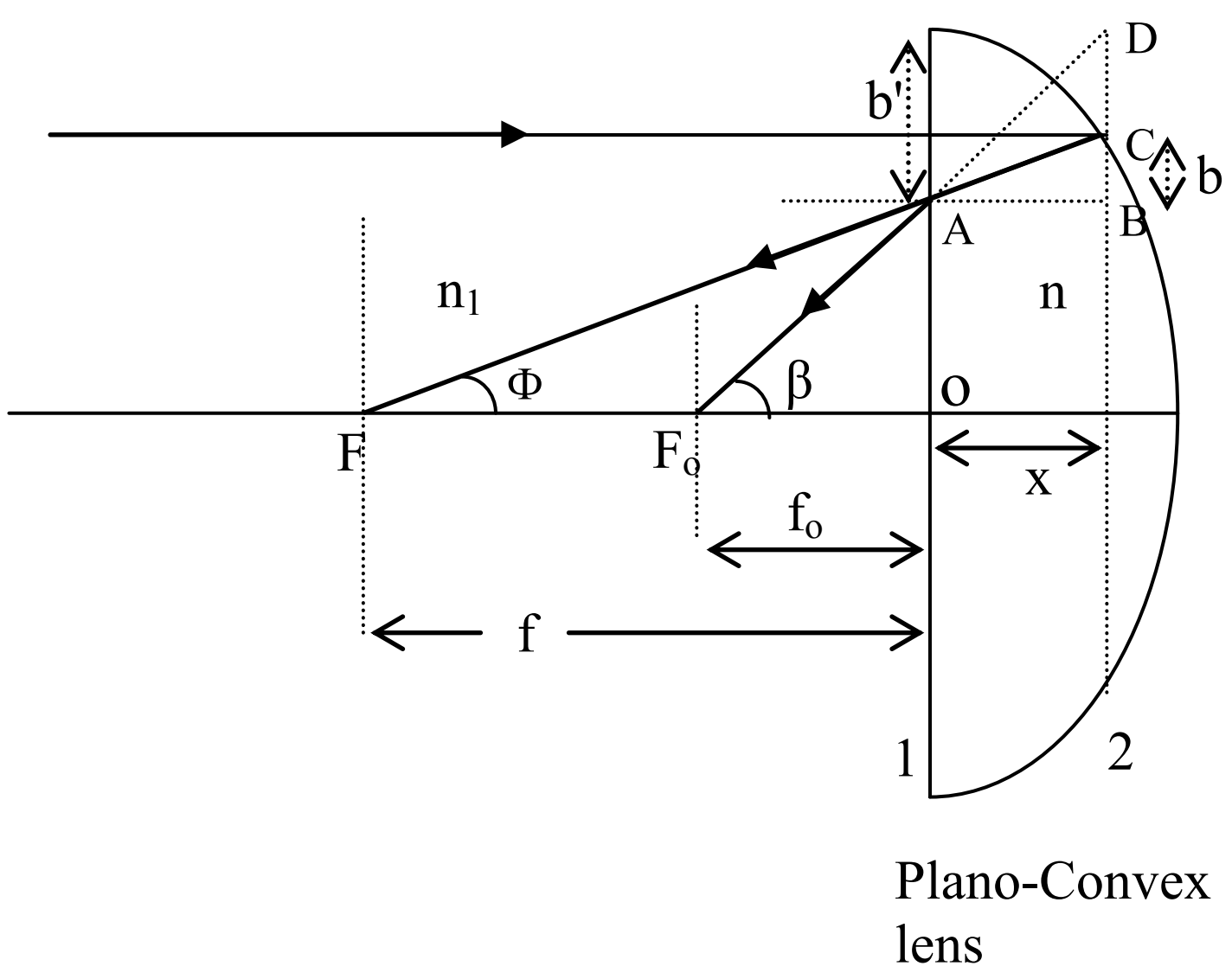

Fig.(2): A collimated beam getting reflected from the plane surface and concave surface of a plano-convex lens resulting in two beams: a parallel beam and a converging beam.
n- refractive index of lens; $n_l$ – refractive index of incident medium.
$A\hat{F}O = C\hat{A}B = \varphi$; $A\hat{F_o}O = D\hat{A}B = \beta$; BC = b; BD = $b'$;
OF = f; $OF_o = f_o$

A parallel beam, after reflection, focuses at a distance, f, in the absence of refraction. But in a plano-convex lens, as a consequence of refraction at the emerging plane surface1, the reflected beam from the surface 2, focuses at a distance $f_o$. From the figure 2, using Snell's law of refraction,

$$n \sin(\Phi) = n_l \sin(\beta)$$

For air medium, $n_l = 1$. Thus

$$n \sin(\Phi) = \sin(\beta) \qquad \text{----} \quad (3)$$

From figure 2, we see that the reflected beam from concave surface should have focused to point F, but due to refraction at surface 1, it actually focuses to $F_o$. If Φ and β are the angles made by the focusing beams at F and $F_o$, then

$$BC = AB \tan(\Phi) \text{ or } \tan(\Phi) = \frac{b}{x} \approx \sin(\Phi) \qquad \text{----} \quad (4)$$

$$\& \ BD = AB \tan(\beta) \text{ or } \tan(\beta) = \frac{b'}{x} \approx \sin(\beta) \qquad \text{----} \quad (5)$$

Therefore,
Eqn (3) becomes

$$b' = n\,b \qquad \text{----} \quad (6)$$

Again from the two triangles, FOA and $F_oOA$,

$$OA = f \tan(\Phi) = f_o \tan(\beta) \qquad \text{----} \quad (7)$$

Using small angle approximation and eqns. 2, 3, 4, 5, we obtain

$$f_o = \frac{f}{n} \qquad \text{----} \quad (8)$$

But for an ordinary concave mirror without any refraction after the reflection, we know that

$$f = \frac{R}{2}$$

Now in the present situation, the reflected beam undergoes refraction and hence the relation is modified by expressing f in terms of $f_o$:

$$R = 2\,n\,f_o \qquad \text{----} \quad (9)$$

This is a very useful relation[5] in this experiment since $f_o$ is directly measured in this setup. Therefore the radius of curvature of plano-convex lens which is generally measured using a spherometer, can now be determined using eqn.(9).

## 4. Derivation of general formula

In the conventional Newton's rings, the two beams which give rise to interference fringes are derived from a single beam as shown in the figure (3a). Whereas in the modified Newton's rings the two rays which form the rings originate from two sides of the optical axis passing through the center of the lens as shown in figure (3b). Denoting the path lengths travelled by the two rays as $P_1$ and $P_2$ where $P_1$ is approximated to be A→B→E→B→C and $P_2$ is C→D→C, the path lengths travelled by the two reflected rays in figure 4 are given by

$$P_1 = x + 2(t\text{-}a)n + \sqrt{x^2 + (r_m + b)^2} \qquad \text{------} \qquad (10)$$

$$P_2 = 2x + \frac{\lambda}{2} \text{ ( due to phase shift of } \pi) \qquad \text{-------} \qquad (11)$$

where n is refractive index and t, the maximum thickness of the lens at the centre.

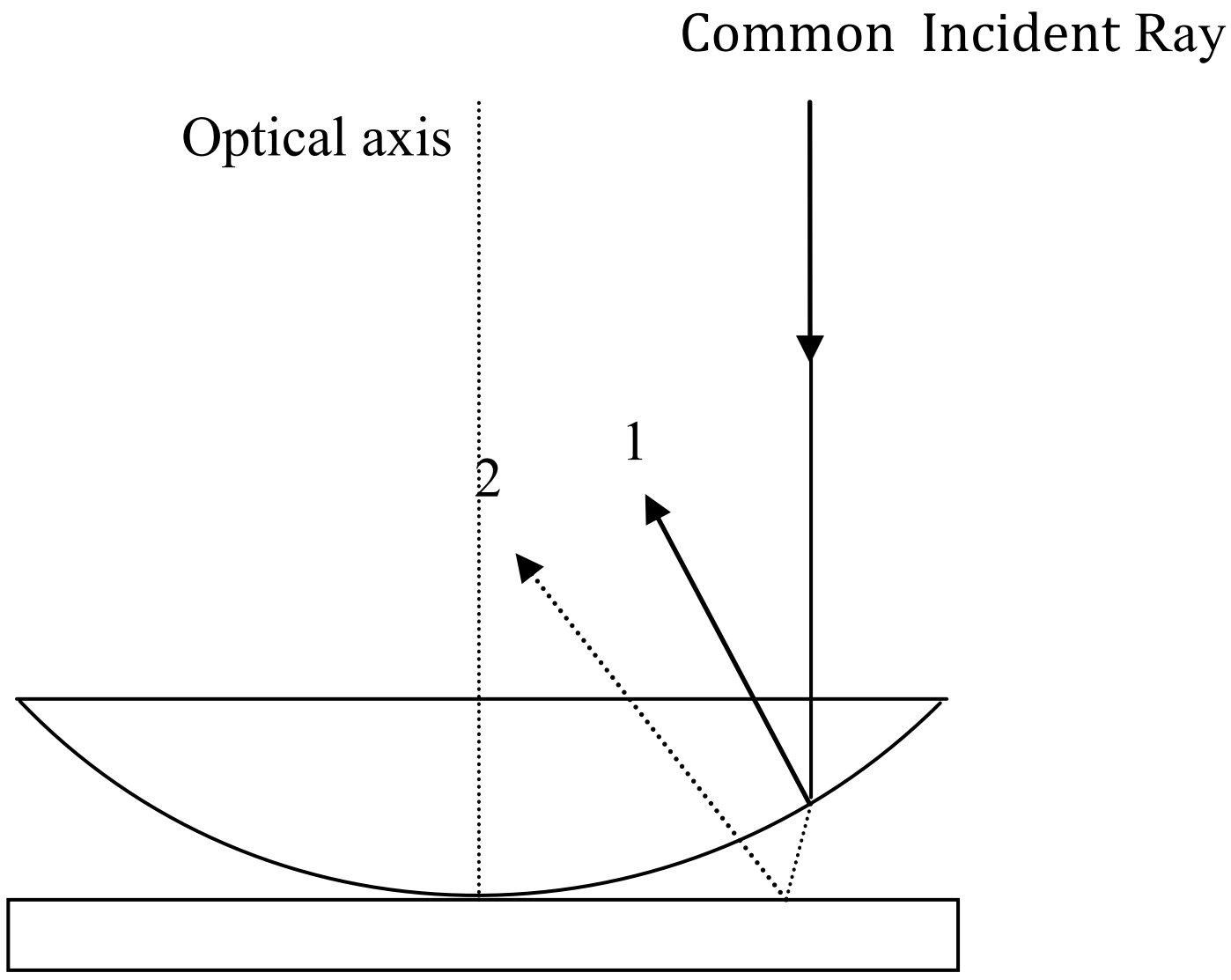

Fig(3a): The two rays which give rise to conventional Newton's rings from a common incident ray are shown above.

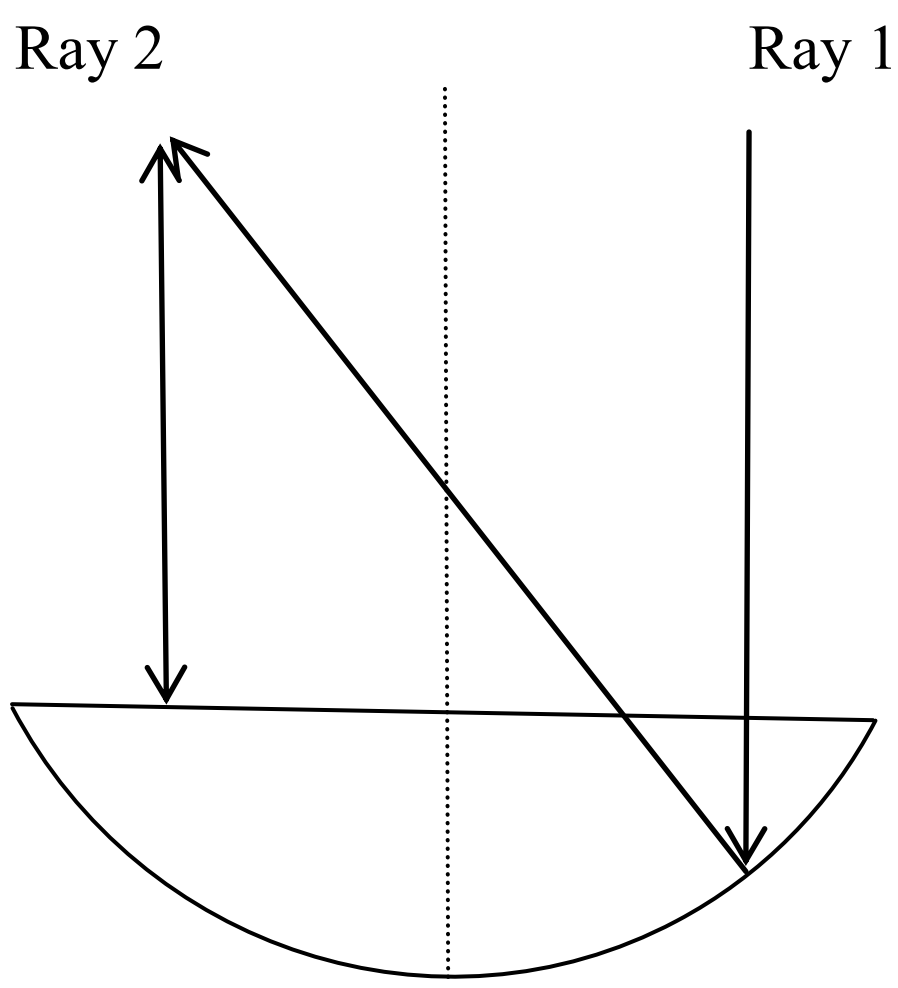

Fig(3b) : The two rays which give rise interference fringes in modified Newton's rings. Ray1 is reflected from the concave surface. Ray2 is reflected from the plane surface

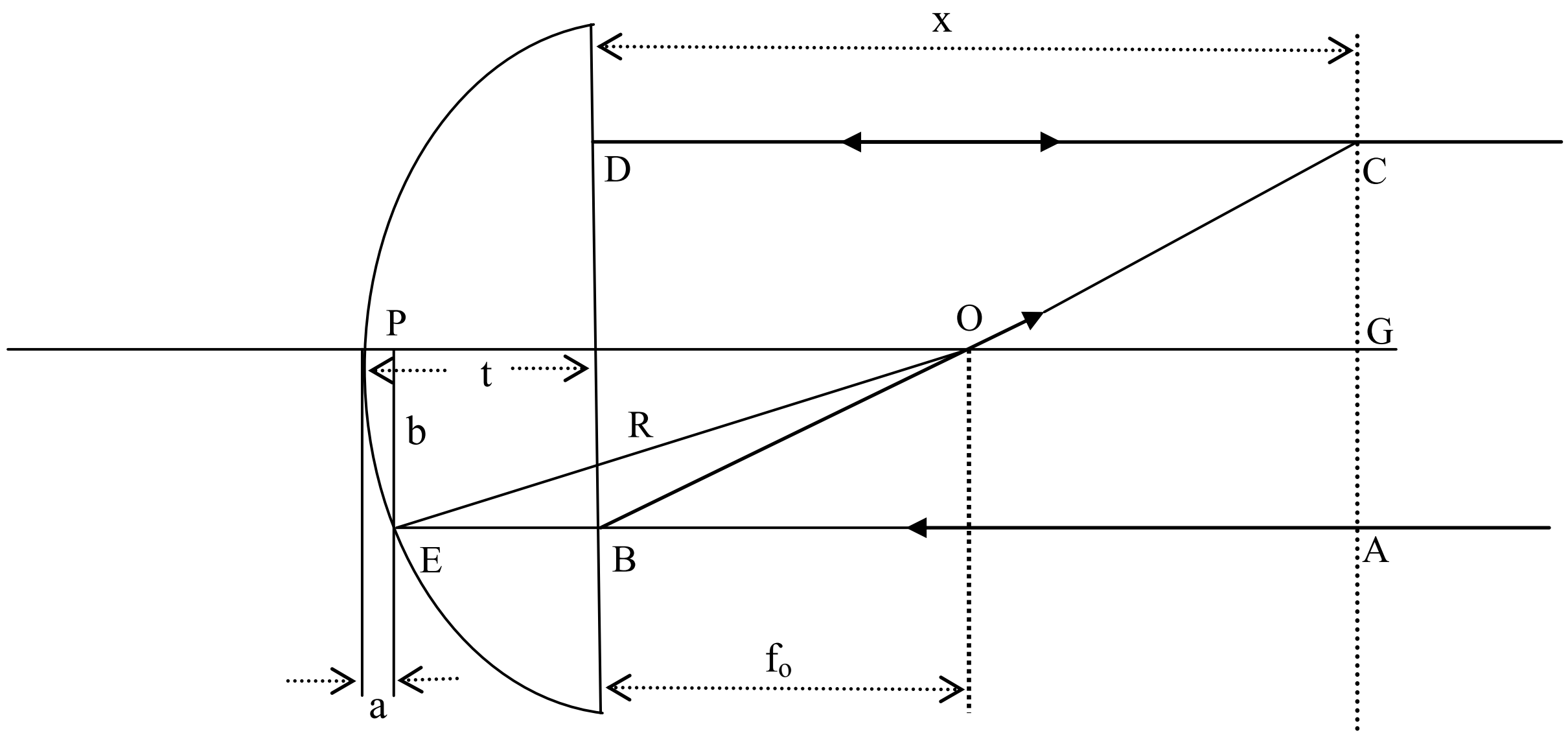

Fig (4): Ray diagram for modified Newton's ring derivation. EO = R, Radius of curvature. CGA : Observation plane, t – maximum thickness of lens, b – off-axis distance ray 1 incident along AB, GC – radius of a ring of m[th] order.

For the ray is incident at the center of the lens along the optical axis, the off-axis distance distance, b=0 and GC= $r_m$ = 0 = a. Hence we get

$$P_1 - P_2 = 2nt - \frac{\lambda}{2}$$

For dark fringes, the path difference should be a multiple of $\frac{\lambda}{2}$. Hence,

$$P_1 - P_2 = \left(N + \frac{1}{2}\right)\lambda$$

N, being the maximum order of the central fringe. Therefore

$$2nt - \frac{\lambda}{2} = \left(N + \frac{1}{2}\right)\lambda$$

or $$2nt = (N+1)\,\lambda$$

In general, rays are incident not only along the optical axis, but also at various off-axis distances, forming a plane wavefront. Therefore, for a ray incident at an off-axis distance 'b' from the optical axis, destructive interference is observed when

$$P_1 - P_2 = M\lambda + \frac{\lambda}{2} \qquad \text{--------} \qquad (12)$$

where M is the order of the dark ring at a distance b from the centre.

$$\left[2nt - 2na + \sqrt{x^2 + (r_m + b)^2}\right] - \left[x + \frac{\lambda}{2}\right] = M\lambda + \frac{\lambda}{2}$$

$$N\lambda + \lambda - 2na + x\sqrt{1 + \left(\frac{r_m + b}{x}\right)^2} - x = M\lambda + \lambda \quad \text{-------} \quad (13)$$

Using the approximation, $\sqrt{1+y} \approx 1 + \frac{y}{2}$ for $y \ll 1$, equation (13) reduces to

$$-2na + x\left[1 + \frac{(r_m + b)^2}{2x^2}\right] - x = m\lambda \qquad \text{--------} \qquad (14)$$

where $m = |M - N|$. With this definition, when M=N, m=0 for the central fringe and increases for outer rings. Thus the ordering of rings becomes identical to conventional Newton's rings.

$$\frac{(r_m + b)^2}{2x} - 2na = m\lambda$$

From the right angle triangle, EOP in figure 3:

$$R^2 = b^2 + (OP)^2$$

$$= b^2 + (R-a)^2 = b^2 + R^2 + a^2 - 2aR$$

Neglecting $a^2$ which is very small, we get the relation $a = \frac{b^2}{2R}$

$$\frac{(r_m + b)^2}{2x} - \frac{2nb^2}{2R} = m\lambda \qquad \text{--------} \qquad (15)$$

We define magnification factor of the fringes as

$$F = \frac{r_m}{b} \qquad \text{--------} \qquad (16)$$

such that at UMP(Unit Magnification Plane), $r_m = b$ and hence F =1.

From the geometry of figure 4, $\frac{r_m}{b} = \frac{x - f_o}{f_o}$. But $R = 2nf_o$, hence

$$F = \frac{x - f_o}{f_o} = \frac{x}{f_o} - 1$$

$$F + 1 = \frac{x}{f_o} \qquad \text{--------} \qquad (17)$$

Using eqn.(16) to substitute for 'b' into eqn.(15), we get

$$\frac{\left(r_m + \frac{r_m}{F}\right)^2}{2x} - \frac{n}{R}\frac{r_m^2}{F^2} = m\lambda$$

$$\frac{r_m^2}{2x}\left(\frac{F+1}{F}\right)^2 - \frac{n}{R}\frac{r_m^2}{F^2} = m\lambda$$

Substituting for `x' from equation (17), we get

$$\frac{r_m^2}{F^2}\left[\frac{(F+1)^2}{2(F+1)f_o} - \frac{n}{R}\right] = m\lambda$$

$$\frac{r_m^2}{F^2}\left[\frac{(F+1)}{2f_o} - \frac{n}{R}\right] = m\lambda$$

But R = 2nf$_o$

$$\frac{r_m^2}{F^2}\left[\frac{(F+1)n}{R} - \frac{n}{R}\right] = m\lambda$$

$$\frac{r_m^2}{F^2}\left[\frac{nF}{R}\right] = m\lambda$$

$$\frac{n\,r_m^2}{F\,R} = m\lambda \qquad \text{----- (18)}$$

This expression gives the relation for the radius of m$^{th}$ order ring. Therefore, the radius of (m+p)$^{th}$ ring is obtained by replacing m by m+p.

$$\frac{n\;r_{m+p}^2}{F\,R} = (m+p)\lambda \qquad \text{----- (19)}$$

Therefore, subtracting these two expressions in equations (18) and (19), then we obtain the formula for the wavelength, given by

$$\frac{(r_{m+p}^2 - r_m^2)n}{p\,F\,R} = \lambda \qquad \text{----- (20)}$$

This formula reduces to the original expression when the magnification factor, F = 1 for unit magnification plane(UMP). In fact, this formula can be employed in this experiment irrespective of where the measurements are made. The magnification factor, F is determined by knowing x, the distance of observation plane from the lens and $f_o$ using eqn.(17).

## 5. Tracker Software:

Another innovation that we have introduced in modified Newton's rings, is the use of CCD camera to capture the fringes instead of making measurements using a travelling microscope. Then the fringes are analysed using 'tracker' software. Tracker[6] is a video analysis software package built on the Open Source Physics (OSP) Java framework. Features of this software include object tracking with position, velocity and acceleration overlays and graphs, special effect filters, multiple reference frames, calibration points, line profiles for analysis of spectra and interference patterns, and dynamic particle models. It is designed to be used in introductory college physics labs and lectures.

Tracker is freely downloadable from the net and can be employed for experiments by anyone in the academic community. Usage of this software involves capturing the video of interest with a webcam or CCD detector, which can be later analyzed frame by frame. It has a host of features which makes scientific analysis very easy. The two important features of interest are the Point Mass Analysis and Line Profile.

The Point Mass Analysis is a technique which is generally employed in the study of experiments in Mechanics and Optics which involve tedious measurements. This feature requires tracking of points. By tracking we mean selecting the points of interest in the video frame by pressing the shift and enter keys simultaneously. As we identify points for tracking, various real-time graphs are generated which may be of interest to us.

The Line Profile is another useful feature which shows us in a graph the variation of intensity along a line. In the case of modified Newton's Rings, we generated the intensity variation with respect to distance shown in figure 6, using this line profile feature. Using the data table or graph of intensity versus distance, the diameters of various rings were easily determined. This Line profile, when performed along different diameters of the rings is equivalent to taking multiple readings which can lead to improved accuracy in the measurement.

The calibration is done using a standard wire of known length or diameter. To employ this in the measurement, a video of this wire is taken and its tape measure is assigned the known standard value. The tape is then copied to the video frames which we need to analyze.

## 6. Experimental observations and Results:

While deriving the formula we have claimed that the process of fringe formation or the two rays which form the interference fringes in the conventional Newton's rings and Modified Newton's rings setup are entirely different. This is clearly shown in the figure (3a) and figure (3b). This can be easily verified while performing the experiment as follows. Once the rings are formed after making the necessary adjustments such as careful collimation, placing the CCD camera at the observation plane(may be at UMP or otherwise), the collimated beam, which falls on the lens, can be gradually blocked manually by an opaque screen. It was noticed that the fringes were found to disappear from the periphery on both sides, one by one. As the screen reaches to block one half of the beam, the rings completely vanish. This is a clear indication regarding which of the two beams which give rise to interference fringes in this setup. When the same process of blocking was repeated in the conventional Newton's rings setup, the rings never vanished completely. The fringes were not seen in regions of blocking in the field of view, whereas in the other half, fringes were seen clearly as before.

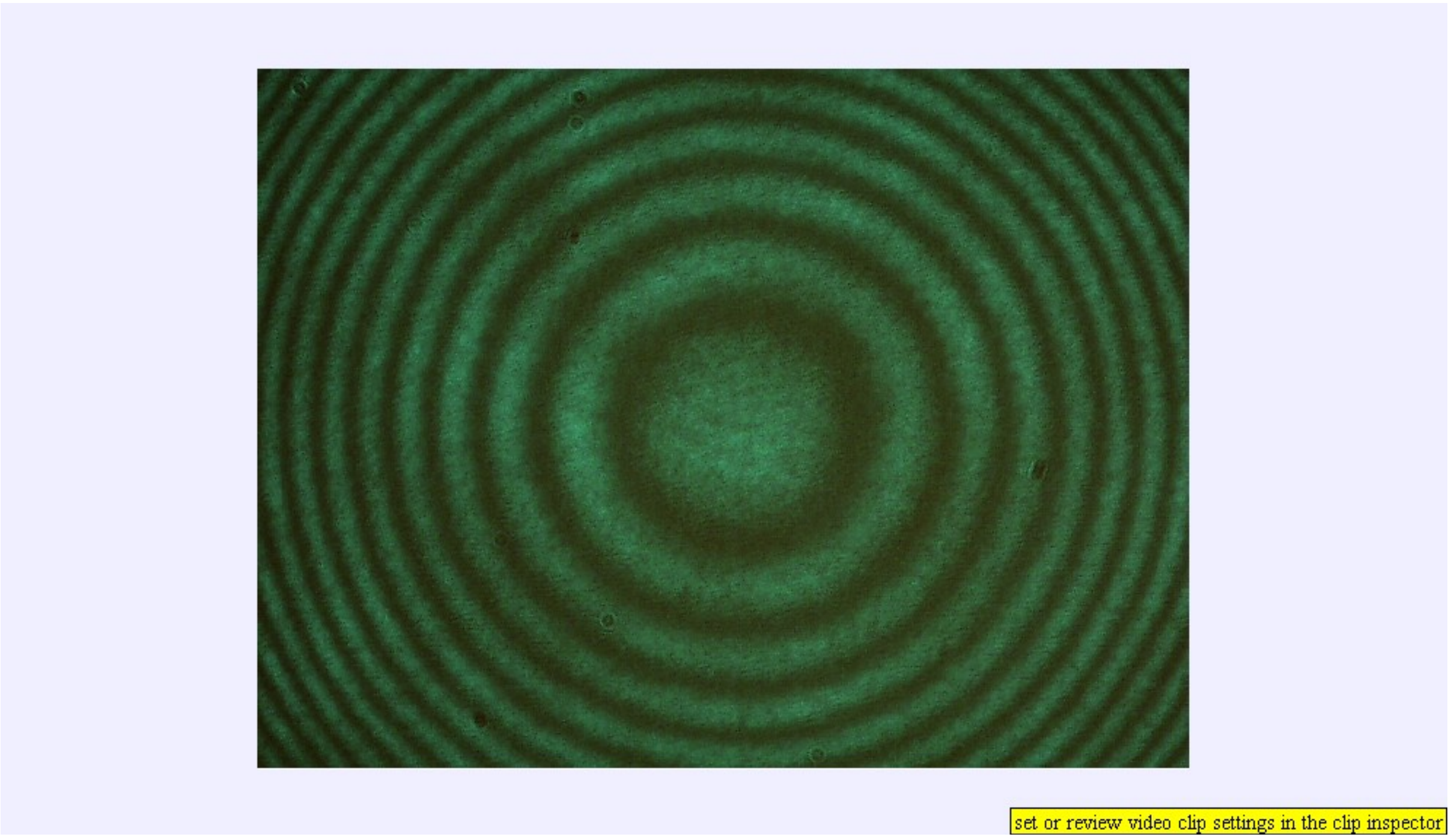

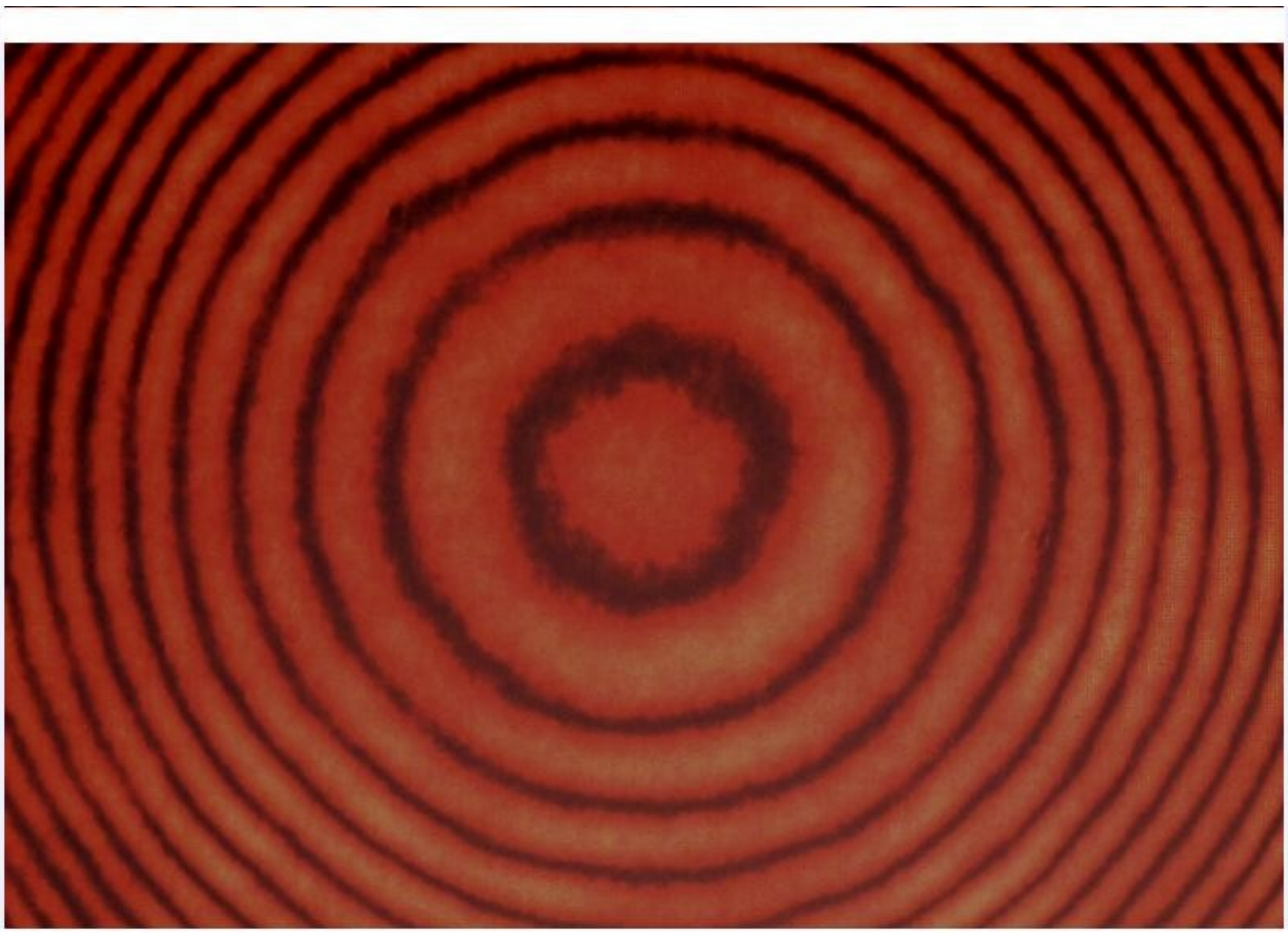

Fig.(5): Typical  Modified Newton's rings using He-Ne laser and Semiconductor green laser

Plotting the data points and the best fit straight line based on linear regression is performed using the software, Scilab. The focal distance for the plano-convex  chosen was $f_o$ = 29cm and hence the radius of curvature R=87cm(=2n$f_o$). The distance x of the observation plane is measured from the plano-convex lens. The unit magnification plane (UMP) is situated at distance of  x = 2$f_o$ =58cm from the plano-convex lens. The wavelength is calculated using the slope of the straight line graph of m versus $r_m^2$, using λ = (slope)n/FR. The magnification factor, F is calculated using eqn.(17). A series of experiments was performed using two different lasers, with observation planes before, after and at the unit magnification plane(UMP). Measurements were made using a travelling microscope as well as using tracker software upon capturing the fringes using a CCD camera. Results of these are given below in the tables and graphs. In the tables, x represents the distance of observation from the plano-convex lens and F is the magnification factor

| Table 1: x = 44cm Slope= .0021cm$^2$ F=0.52 λ = 625nm (Travelling Microscope) | | | | | |
|---|---|---|---|---|---|
| Ring no. | LHS | RHS | Diameter cm | Radius R cm | $R^2$ cm$^2$ |
| 5 | 9.539 | 9.350 | 0.189 | 0.0945 | 0.0089 |
| 4 | 9.529 | 9.363 | 0.166 | 0.0830 | 0.0069 |
| 3 | 9.516 | 9.375 | 0.141 | 0.0705 | 0.0050 |
| 2 | 9.499 | 9.385 | 0.114 | 0.0570 | 0.0032 |
| 1 | 9.479 | 9.450 | 0.029 | 0.0145 | 0.0002 |

| Table 2: x = $2f_o$=58cm Slope= .0038cm$^2$ F=1 λ = 649nm (Travelling Microscope) | | | | | |
|---|---|---|---|---|---|
| Ring no. | LHS | RHS | Diameter cm | Radius R cm | $R^2$ cm$^2$ |
| 5 | 7.997 | 7.732 | 0.265 | 0.1325 | 0.0176 |
| 4 | 7.980 | 7.741 | 0.239 | 0.1195 | 0.0143 |
| 3 | 7.963 | 7.760 | 0.203 | 0.1015 | 0.0103 |
| 2 | 7.943 | 7.780 | 0.163 | 0.0815 | 0.0066 |
| 1 | 7.913 | 7.812 | 0.101 | 0.0505 | 0.0026 |

| Table 3: x = 65.5cm Slope= .0046cm$^2$ F=1.26 λ = 635nm (Travelling Microscope) | | | | | |
|---|---|---|---|---|---|
| Ring no. | LHS | RHS | Diameter cm | Radius R cm | $R^2$ cm$^2$ |
| 5 | 9.124 | 8.830 | 0.294 | 0.1470 | 0.0216 |
| 4 | 9.108 | 8.844 | 0.264 | 0.1320 | 0.0174 |
| 3 | 9.085 | 8.867 | 0.218 | 0.1090 | 0.0119 |
| 2 | 9.067 | 8.888 | 0.179 | 0.0895 | 0.0080 |
| 1 | 9.031 | 8.919 | 0.112 | 0.0560 | 0.0031 |

| Table 4: x = 44cm Slope=0.0019cm$^2$ F=.52 λ = 629nm (Tracker Software) | | | | | |
|---|---|---|---|---|---|
| Ring No | LHS | RHS | Diameter cm | Radius R cm | $R^2$ cm$^2$ |
| 1 | -0.0288 | 0.0293 | 0.0581 | 0.0291 | 0.0008 |
| 2 | -0.0518 | 0.0518 | 0.1036 | 0.0518 | 0.0027 |
| 3 | -0.0669 | 0.0693 | 0.1362 | 0.0681 | 0.0046 |
| 4 | -0.0801 | 0.0814 | 0.1615 | 0.0807 | 0.0065 |
| 5 | -0.0907 | 0.0924 | 0.1831 | 0.0915 | 0.0084 |
| 6 | -0.1011 | 0.1014 | 0.2026 | 0.1013 | 0.0103 |

| Table 5: x = 2$f_o$= 58cm Slope=0.0038cm$^2$ F=1 λ = 650 nm (Tracker Software) | | | | | |
|---|---|---|---|---|---|
| Ring No | LHS | RHS | Diameter cm | Radius R cm | $R^2$ cm$^2$ |
| 1 | -0.0323 | 0.0340 | 0.0663 | 0.0332 | 0.0011 |
| 2 | -0.0729 | 0.0658 | 0.1387 | 0.0693 | 0.0048 |
| 3 | -0.0948 | 0.0893 | 0.1842 | 0.0921 | 0.0085 |
| 4 | -0.1129 | 0.1107 | 0.2236 | 0.1118 | 0.0125 |
| 5 | -0.1283 | 0.1261 | 0.2543 | 0.1272 | 0.0162 |
| 6 | -0.1425 | 0.1398 | 0.2823 | 0.1411 | 0.0199 |
| 7 | -0.1551 | 0.1524 | 0.3075 | 0.1537 | 0.0236 |

| Table 6: x = +2cm Slope=0.0039cm$^2$ F=1.07 λ = 629nm (Tracker Software) | | | | | |
|---|---|---|---|---|---|
| Ring No | LHS | RHS | Diameter cm | Radius R cm | $R^2$ cm$^2$ |
| 1 | 0.0342 | 0.0352 | 0.0694 | 0.0347 | 0.0012 |
| 2 | 0.0702 | 0.0712 | 0.1414 | 0.0707 | 0.0050 |
| 3 | 0.0950 | 0.0962 | 0.1912 | 0.0956 | 0.0091 |
| 4 | 0.1140 | 0.1152 | 0.2292 | 0.1146 | 0.0131 |
| 5 | 0.1300 | 0.1312 | 0.2612 | 0.1306 | 0.0171 |
| 6 | 0.1444 | 0.1458 | 0.2902 | 0.1451 | 0.0211 |
| 7 | 0.1576 | 0.1584 | 0.3160 | 0.1580 | 0.0250 |

| Table 7: x = 44cm Slope=0.0016cm$^2$ F=0.52 λ = 533nm (Tracker Software) | | | | | |
|---|---|---|---|---|---|
| Ring No | LHS | RHS | Diameter cm | Radius R cm | $R^2$ cm$^2$ |
| 1 | -0.0340 | 0.0302 | 0.0642 | 0.0321 | 0.0010 |
| 2 | -0.0524 | 0.0502 | 0.1026 | 0.0513 | 0.0026 |
| 3 | -0.0664 | 0.0637 | 0.1301 | 0.0651 | 0.0042 |
| 4 | -0.0772 | 0.0750 | 0.1523 | 0.0761 | 0.0058 |
| 5 | -0.0869 | 0.0853 | 0.1722 | 0.0861 | 0.0074 |
| 6 | -0.0961 | 0.0939 | 0.1900 | 0.0950 | 0.0090 |
| 7 | -0.1042 | 0.1020 | 0.2062 | 0.1031 | 0.0106 |

| Table 8: x = 2$f_o$=58cm Slope=0.0031cm$^2$ F=1 λ = 534nm (Tracker Software) | | | | | |
|---|---|---|---|---|---|
| Ring No | LHS | RHS | Diameter cm | Radius R cm | $R^2$ cm$^2$ |
| 1 | -0.0323 | 0.0296 | 0.0619 | 0.0310 | 0.0010 |
| 2 | -0.0646 | 0.0619 | 0.1265 | 0.0633 | 0.0040 |
| 3 | -0.0845 | 0.0829 | 0.1674 | 0.0837 | 0.0070 |
| 4 | -0.1012 | 0.1001 | 0.2013 | 0.1007 | 0.0101 |
| 5 | -0.1152 | 0.1141 | 0.2293 | 0.1147 | 0.0131 |
| 6 | -0.1287 | 0.1276 | 0.2563 | 0.1281 | 0.0164 |
| 7 | -0.1394 | 0.1400 | 0.2794 | 0.1397 | 0.0195 |

| Table 9: x = 65cm Slope=0.0039 cm$^2$ F=1.20 λ = 536nm (Tracker Software) | | | | | |
|---|---|---|---|---|---|
| Ring No | LHS | RHS | Diameter cm | Radius R cm | $R^2$ cm$^2$ |
| 1 | -0.0427 | 0.0448 | 0.0875 | 0.0437 | 0.0019 |
| 2 | -0.0756 | 0.0756 | 0.1512 | 0.0756 | 0.0057 |
| 3 | -0.0983 | 0.0977 | 0.1960 | 0.0980 | 0.0096 |
| 4 | -0.1161 | 0.1166 | 0.2327 | 0.1163 | 0.0135 |
| 5 | -0.1312 | 0.1323 | 0.2635 | 0.1317 | 0.0174 |
| 6 | -0.1452 | 0.1458 | 0.2910 | 0.1455 | 0.0212 |
| 7 | -0.1576 | 0.1587 | 0.3164 | 0.1582 | 0.0250 |

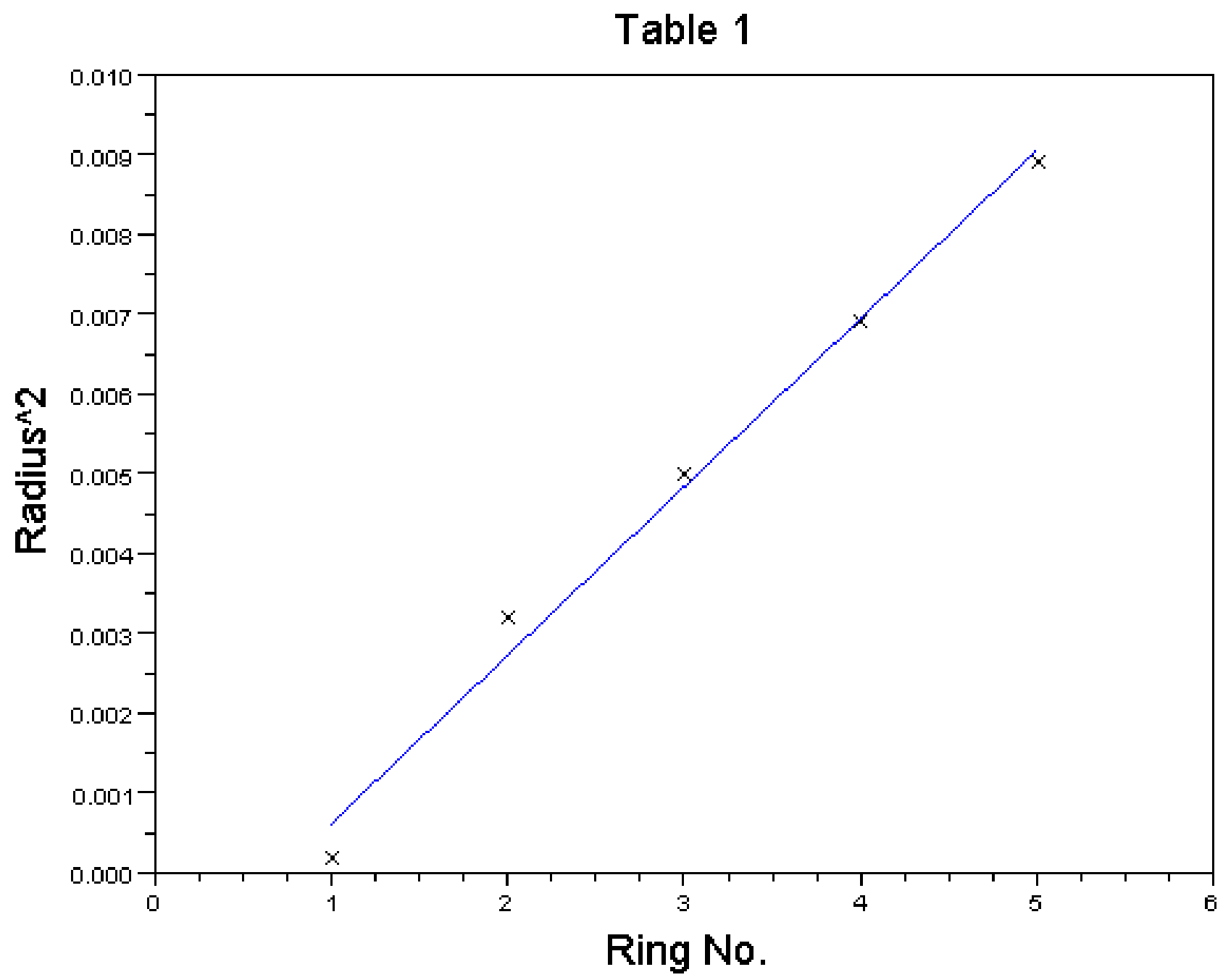
Table 1
Radius^2
Ring No.
0.000
0.001
0.002
0.003
0.004
0.005
0.006
0.007
0.008
0.009
0.010
0
1
2
3
4
5
6

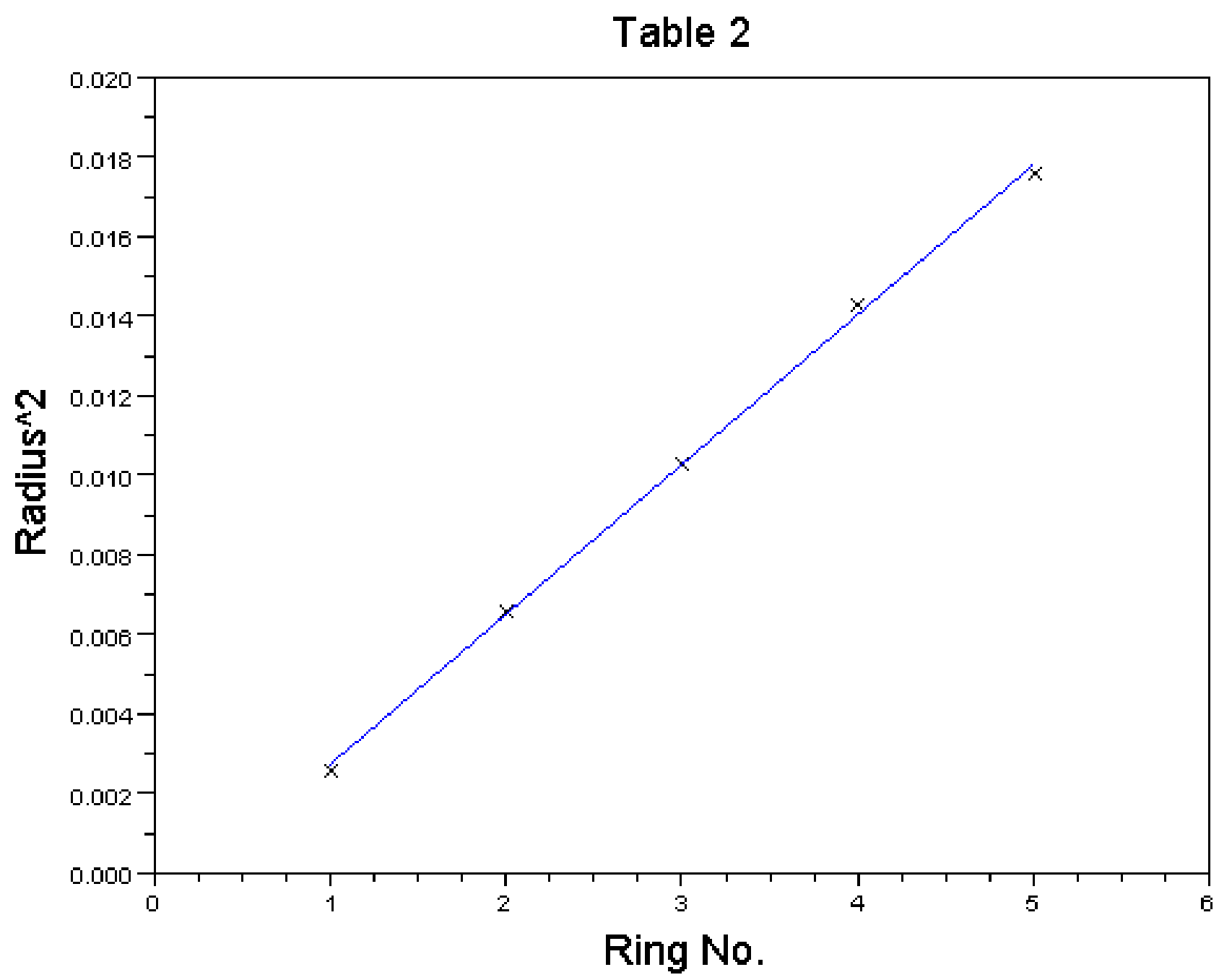
Table 2
Radius^2
Ring No.
0.000
0.002
0.004
0.006
0.008
0.010
0.012
0.014
0.016
0.018
0.020
0
1
2
3
4
5
6

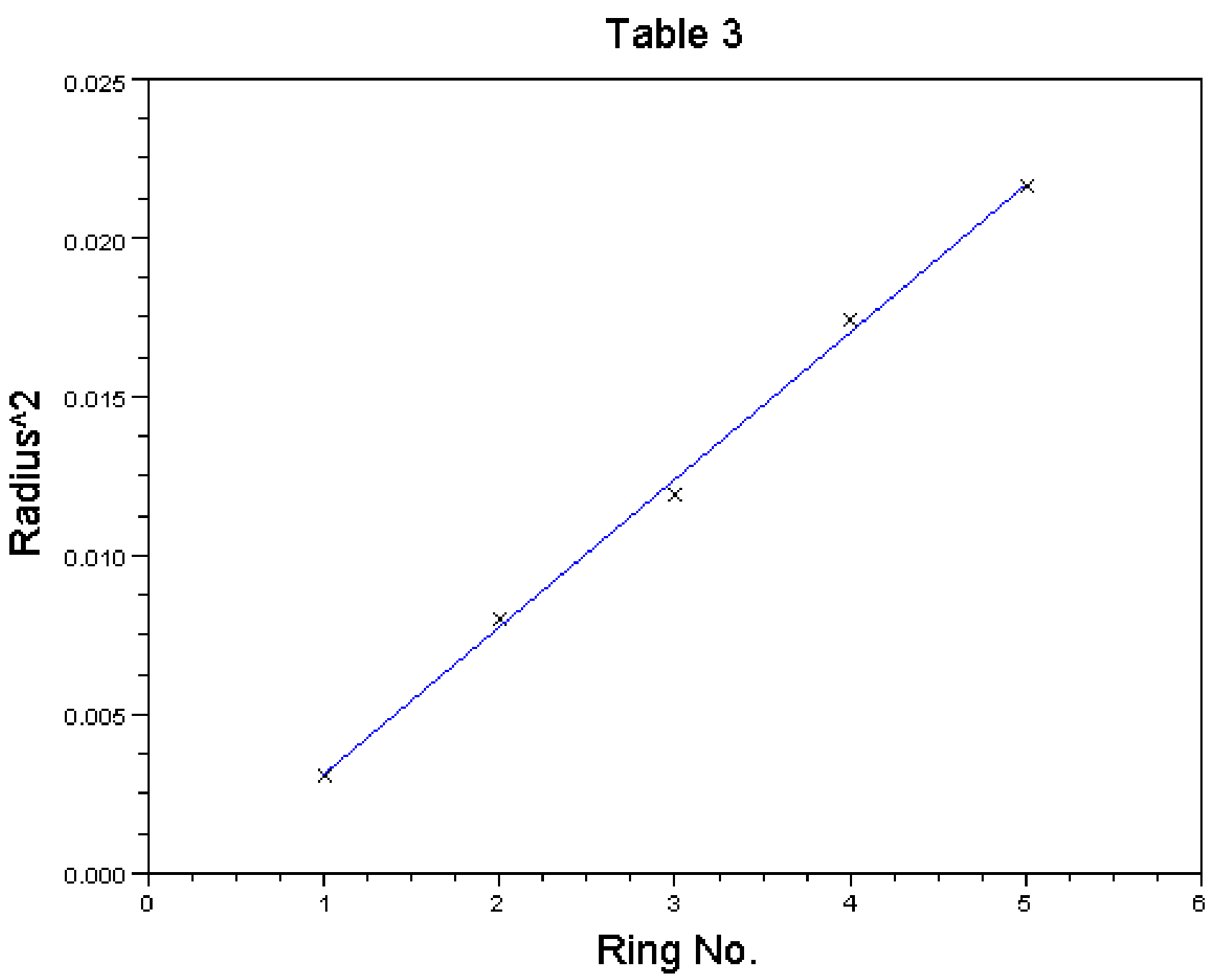

Table 3
0.025
0.020
0.015
0.010
0.005
0.000
Radius^2
0
1
2
3
4
5
6
Ring No.

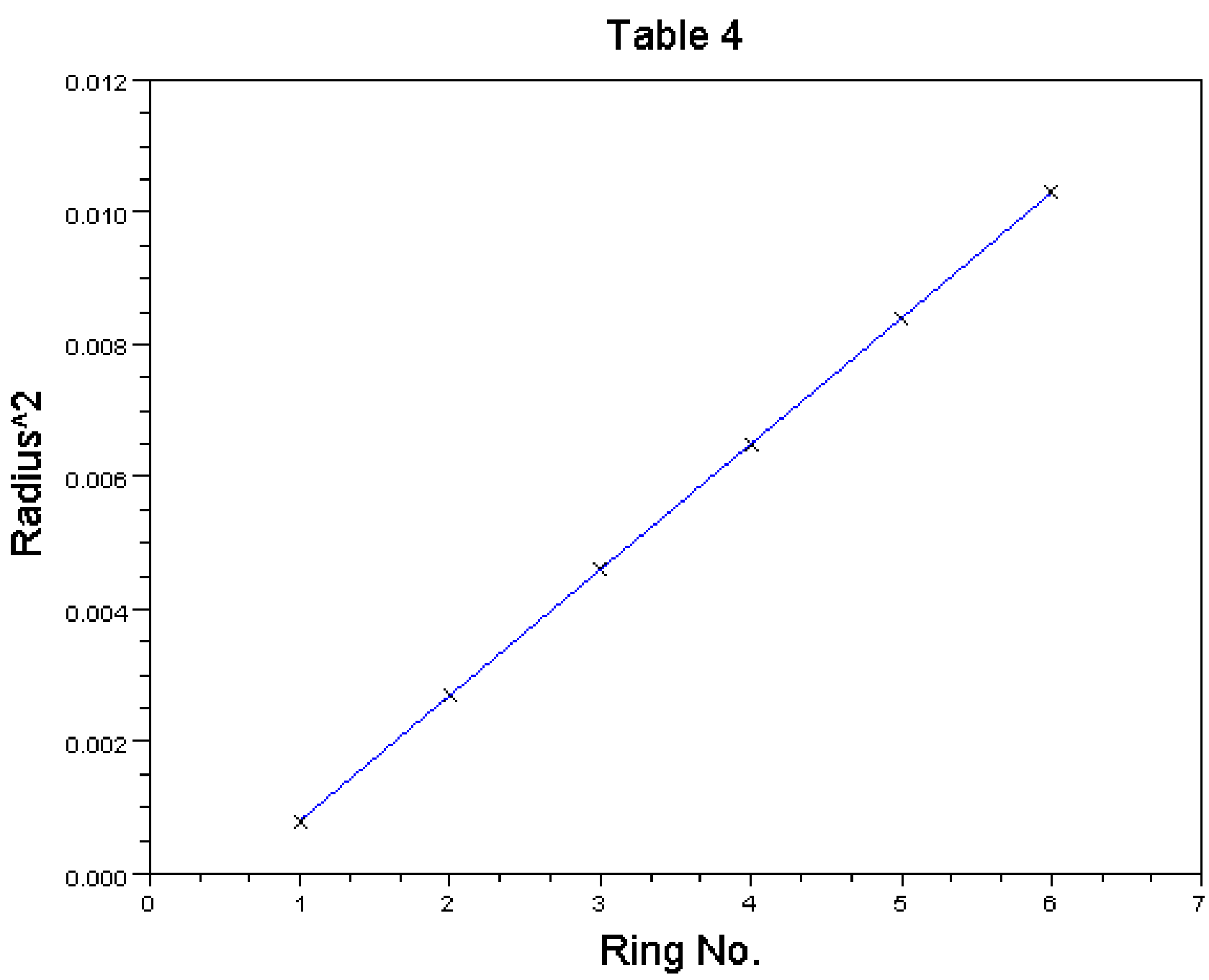

Table 4
0.012
0.010
0.008
0.006
0.004
0.002
0.000
Radius^2
0
1
2
3
4
5
6
7
Ring No.

Table 5

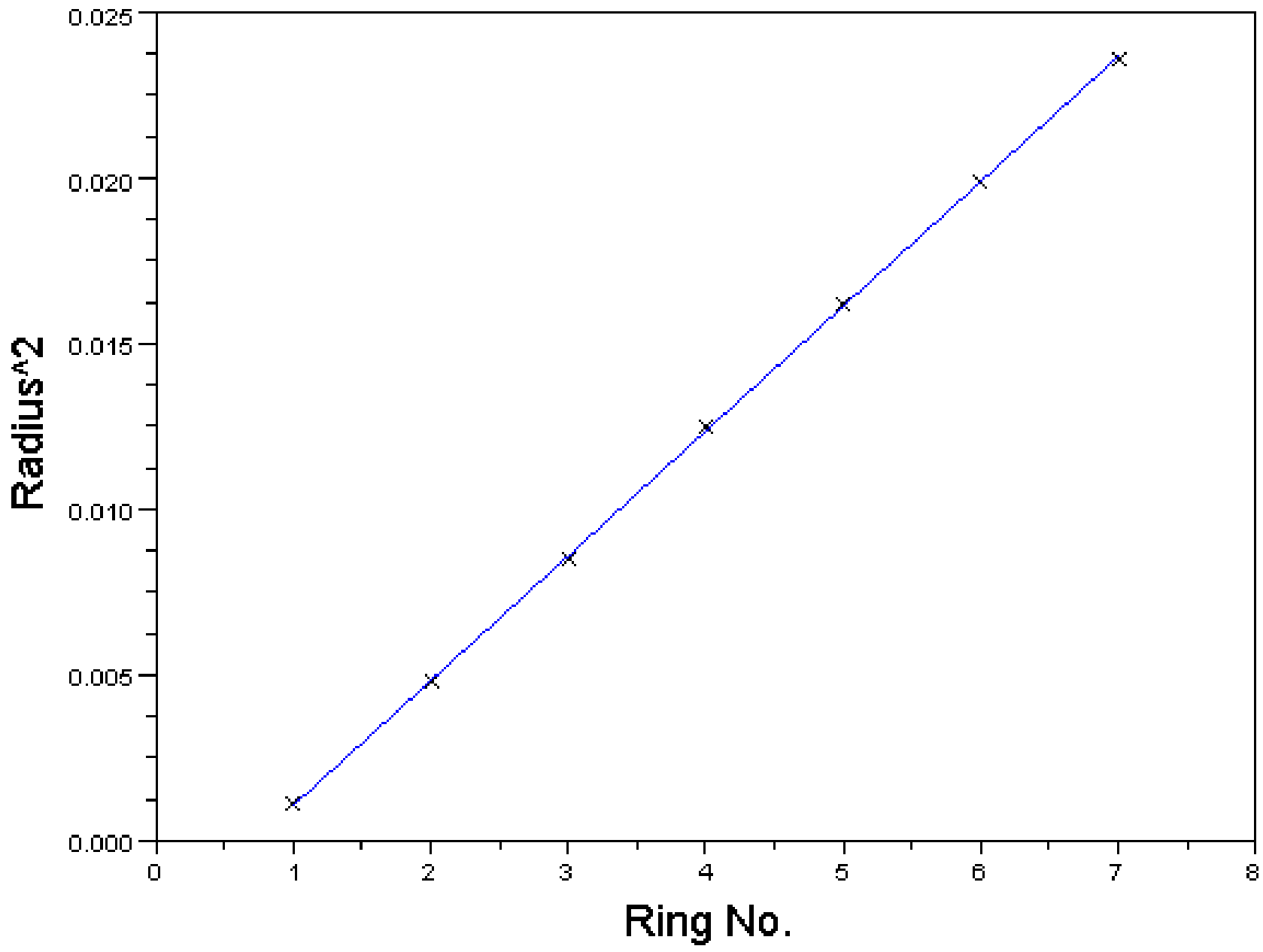

Table 6

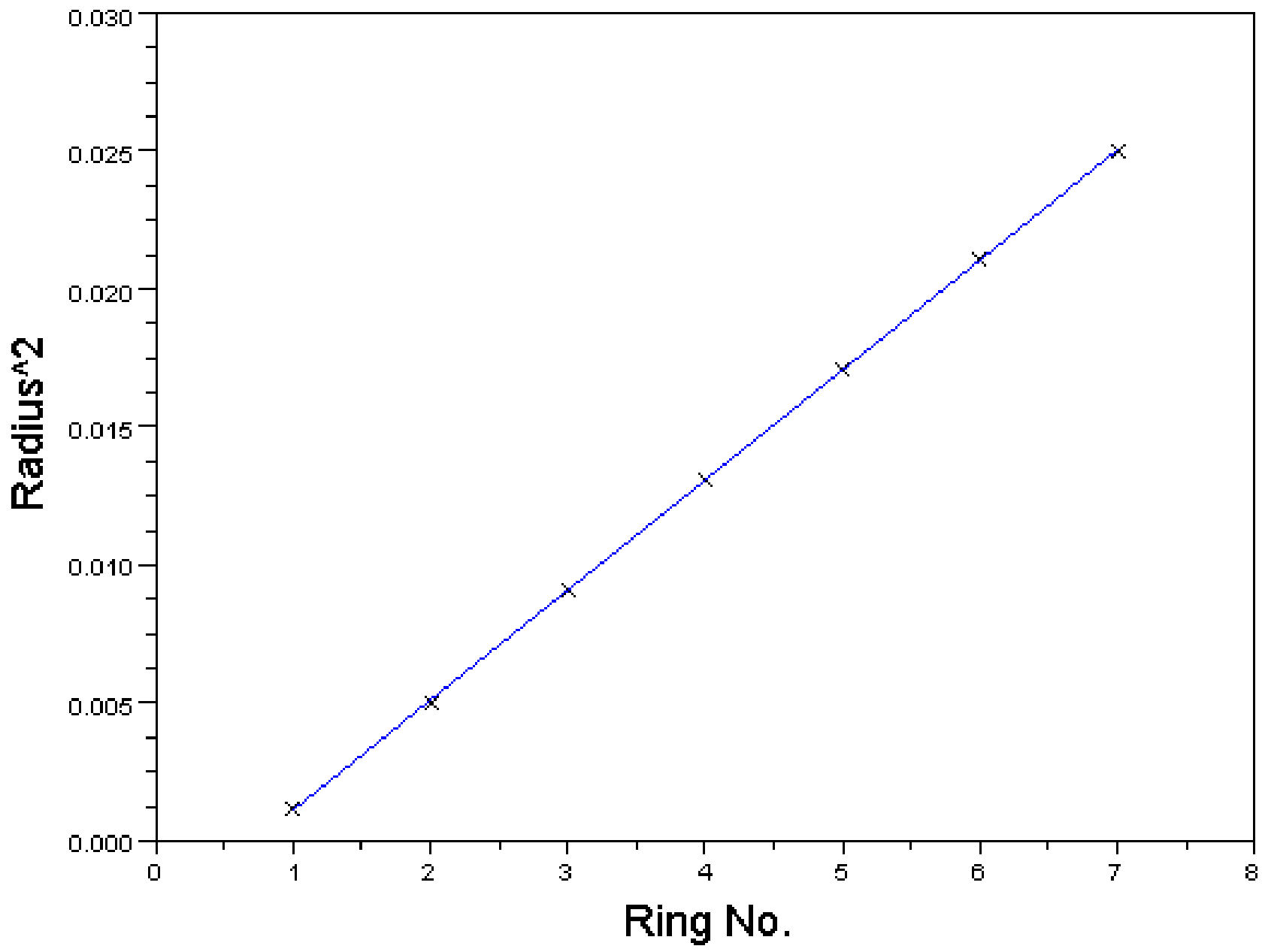

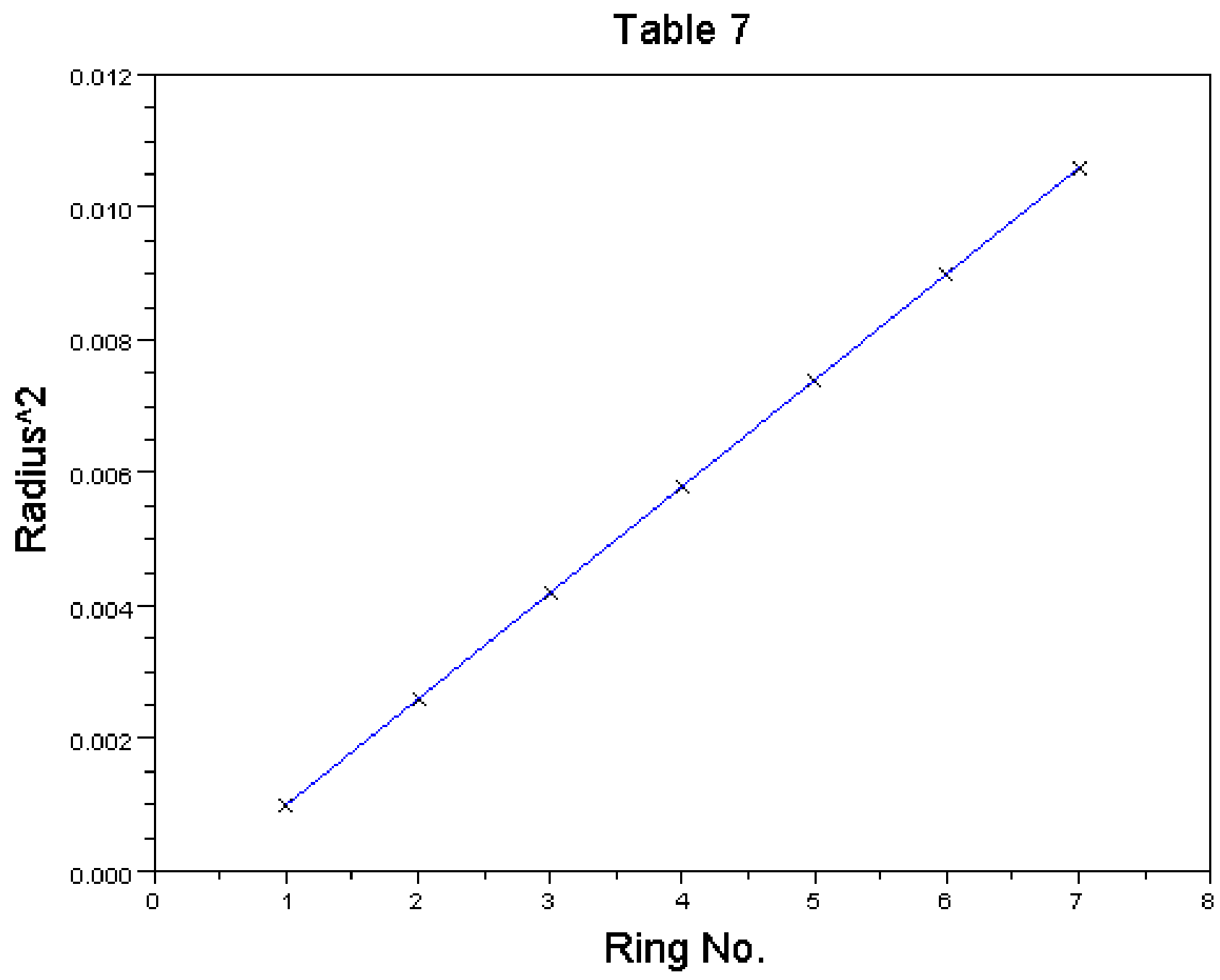
Table 7
Radius^2
Ring No.
0.000
0.002
0.004
0.006
0.008
0.010
0.012
0
1
2
3
4
5
6
7
8

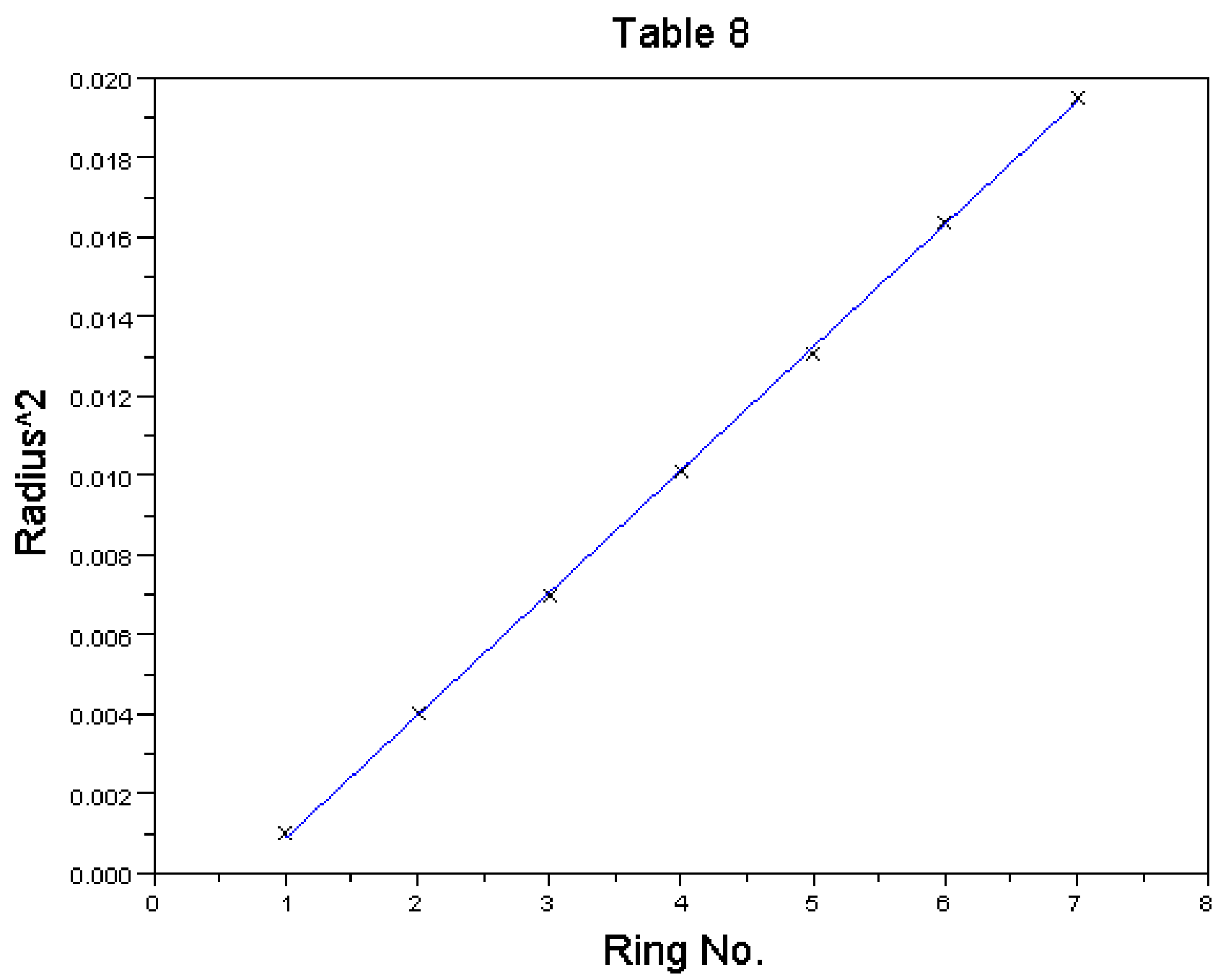
Table 8
Radius^2
Ring No.
0.000
0.002
0.004
0.006
0.008
0.010
0.012
0.014
0.016
0.018
0.020
0
1
2
3
4
5
6
7
8

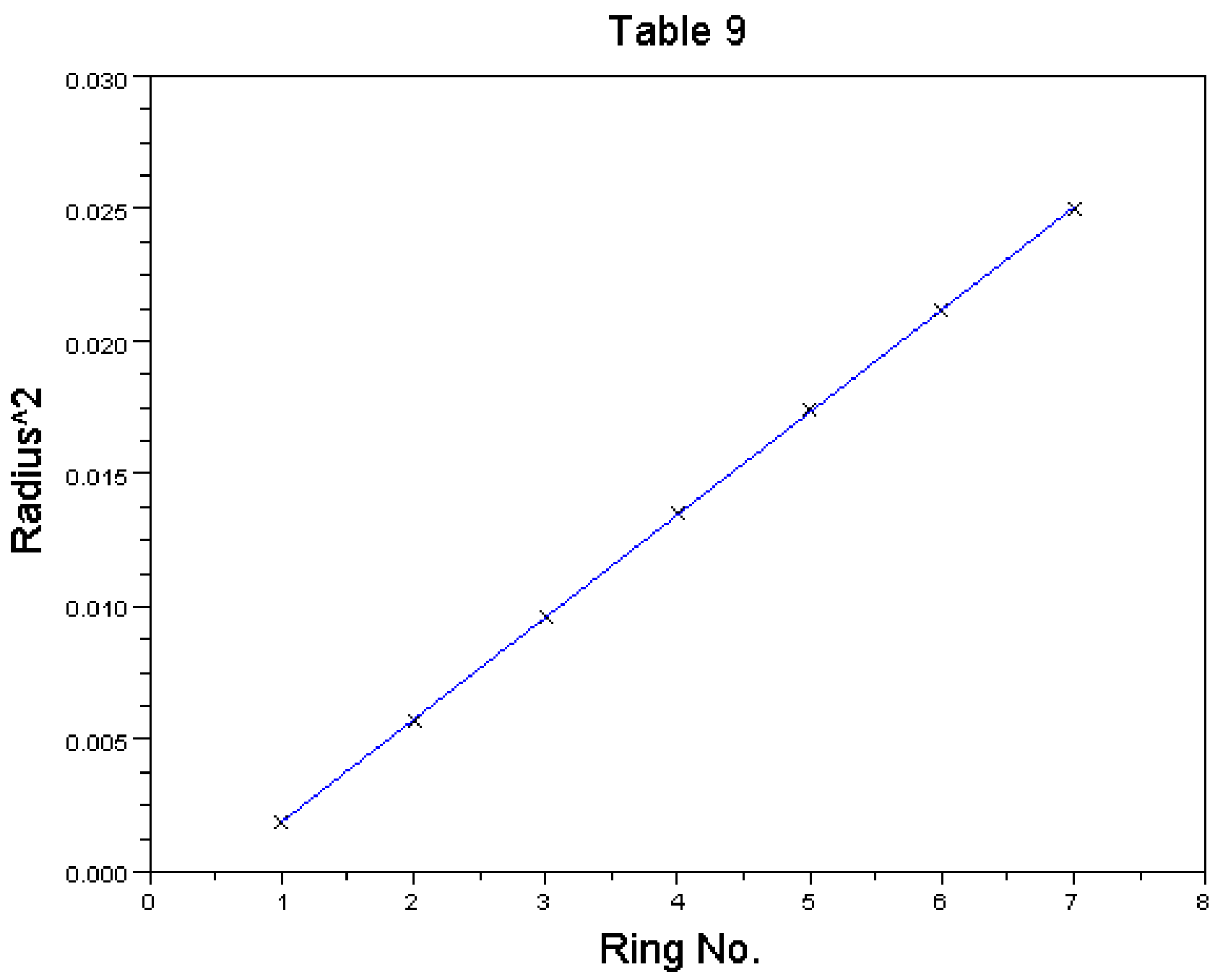

## 7. Estimation of accuracy:

The accuracy or the error in the determination of wavelength can be estimated using the following formula.

$$\frac{\delta\lambda}{\lambda} = \frac{\delta(r_{m+p}^2 - r_m^2)}{r_{m+p}^2 - r_m^2} + \frac{\delta f_o}{f_o} + \frac{\delta x}{x}$$

Accuracy in our measurements varied from 3% to 7%. The first term in the above expression represents the percentage error in the slope of the best fit straight line graph[7]. Second term denotes the percentage error in $f_o$. It was found that when the measurements are made at unit magnification plane, the error was minimum as expected. This is because the last term in the above equation will be absent since at UMP, x=0. Moreover, the fringes will have maximum contrast at UMP and the contrast reduces as one goes away from the unit magnification plane.

| Summary of Results | | | |
|---|---|---|---|
| Table | Source | Analysis Using | Expected λ |
| 1, 2, 3 | He-Ne laser | Travelling microscope | 633 nm |
| 4, 5, 6 | He-Ne laser | Tracker Software | 633 nm |
| 7, 8, 9 | Semicond.laser | Tracker Software | 532 nm |

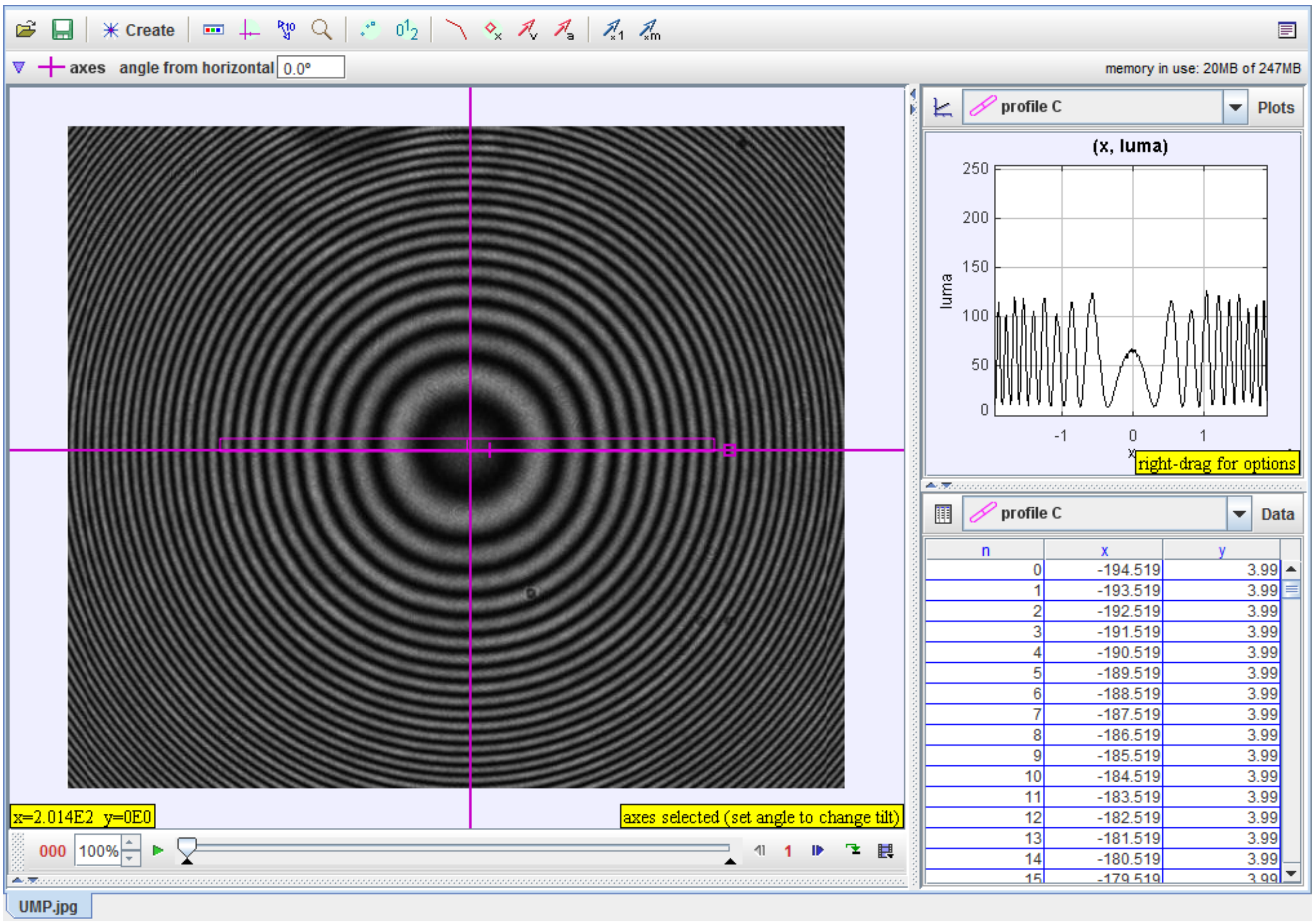

Fig(6): Line Profile analysis of modified Newton's rings using Tracker software. On the right hand side are a graph of distance versus intensity and a table of data which are generated by tracker

## 8. Conclusion:

General formula for wavelength has been derived for modified Newton's rings experiment, which is valid for observation planes at any distance. Experimental data is provided which prove the validity of this formula. Modified Newton's rings experiment has been made simpler by the use of Tracker software by capturing the fringes using a CCD camera. This does away with tedious measurements using a travelling microscope and the analysis is done with more ease and sophistication using the tracker software. Yet another simplification that has been achieved is in the measurement of radius of curvature, R. By knowing $f_o$, the focal distance which is usually measured in this experiment, the radius of curvature is determined by $R = 2nf_o$. Since this involves measuring only single quantity, $f_o$ unlike the method using spherometer which requires two parameters, the error gets reduced to that extent. A simple experimentation revealed the clear distinction between conventional Newton's rings and modified Newton's rings regarding the rays which give rise to interference fringes in the two cases.

## Acknowledgements

Authors gratefully thank Dr. Debendranath Sahoo for going through the manuscript and giving valuable suggestions to improve the presentation of this article.